# New Science on The Open Science Grid

The Open Science Grid Executive Board on behalf of the OSG Consortium:

**Ruth Pordes[1], Mine Altunay[1], Paul Avery[2], Alina Bejan[3], Kent Blackburn[4], Alan Blatecky[5], Rob Gardner[3], Bill Kramer[6], Miron Livny[7], John McGee[5], Maxim Potekhin[8], Rob Quick[8], Doug Olson[6], Alain Roy[6], Chander Sehgal[1], Torre Wenaus[8], Mike Wilde[3], Frank Würthwein[10]**

[1]Fermi National Accelerator Laboratory, [2]University of Florida, [3]University of Chicago, [4]California Institute of Technology, [5]Renaissance Computing Institute [6]Lawrence Berkeley National Laboratory, [7]University of Wisconsin, Madison, [8]Indiana University, [9]Brookhaven National Laboratory, [10]University of California, San Diego

Corresponding Author: ruth@fnal.gov[1]

**Abstract**: The Open Science Grid (OSG) includes work to enable new science, new scientists, and new modalities in support of computationally based research. There are frequently significant sociological and organizational changes required in transformation from the existing to the new. OSG leverages its' deliverables to the large scale physics experiment member communities to benefit new communities at all scales through activities in education, engagement and the distributed facility. As a partner to the poster and tutorial at SciDAC 2008, this paper gives both a brief general description and some specific examples of new science enabled on the OSG. More information is available at the OSG web site: www.opensciencegrid.org.

## 1. Science on the OSG

The Open Science Grid (OSG)[1] provides a nationally distributed computational infrastructure for science research where computing and storage owned by members of an open consortium are made accessible to and shared with other members through installation of OSG software and subscription to OSG services. The majority of the resources are from and current use is by the large physics collaborations, ATLAS, CDF, CMS, D0 and LIGO. This new way of doing their science has already paved the way for a substantially expanded infrastructure, which now includes access to the computing and storage facilities at four DOE laboratories, more than forty universities from twenty states in the US, as well as six sites offshore, and diversified use of the facility to include more than ten other scientific domains (see Figure 1).

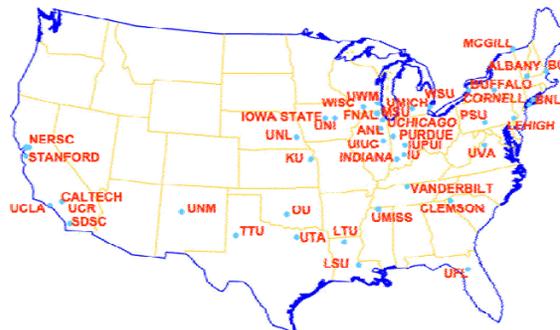

**Figure 1: Sites on the OSG, June 2008**

[1] Fermilab is supported by the U.S. Department of Energy under contract No. DE-AC02-76CH03000.

The mission of the OSG includes extending the shared computational environment to include new participants – new scientists, new campuses, new communities, and new scales – and to ultimately transform the landscape for future research and discovery. Such a transformation is neither light nor fast. It takes months or years of continual communication and support to achieve the cultural as well as the technical adaptation to move an individual and/or a community group from an application environment that assumes dedicated local processing and data repositories to a distributed system of a dynamic set of heterogeneous remote computing and data resources. This implies moving from a local environment characterized by deterministic response times and interactive access to the job and data file execution nodes to a distributed system where performance is achieved via the aggregate overall throughput, where a fraction of the resources are unavailable (inaccessible, preoccupied) for use at any particular time, where the data must be transferred from and to remote processing units, and the resources are shared subject to autonomous local policies of prioritization and access.

The OSG facility provides support, operational security, common software, and other grid-wide facility services such as accounting, monitoring and resource information on which all OSG members rely[2]. For the LHC experiments these OSG services form part of the "critical service" baseline of the World Wide LHC Computing Grid[3]. OSG is thus part of the federated global infrastructure enabling science across more than thirty-five countries on four continents. The facility provides monitoring, management, response and problem solving frameworks which leverage available expertise, promote common best practices and approaches and provide natural dissemination and growth of the knowledge base.

Another important aspect of the OSG is the provision and support of a common, integrated, supported set of software components, the Virtual Data Toolkit (VDT)[4], available on many platforms for installation and use by resources providers and users of the infrastructure. By being available for many variants of Linux (and in client mode for MacOSX and AIX), the VDT provides an anchor to define common interfaces to the services (job execution, data movement and storage, security, management and monitoring etc) on and through the OSG. The components of the VDT are determined by the needs of the communities, and the processes to build, test, and release the software ensure managed evolution and extension of the capabilities offered.

It is eighteen months since the start of the SciDAC-2/NSF funded OSG project. We describe here the types and examples of the new science enabled through the work of the OSG community.

**2. New Scientists**
OSG education and engagement activities work to help and teach new scientists – and, in the case of education, specifically young new scientists – to adapt and run data and compute intensive applications on the existing distributed infrastructure.

OSG grid schools, lasting one to several days in duration, have combined lectures and hands-on laboratories to teach the fundamentals of grid technologies. To date, hey have enabled more than three hundred students to actually run jobs across and transport data between specific OSG sites that advertise support for the "education community". Selected students also attend the longer, residential school, the International Summer School on Grid Computing '08, now co-sponsored by the OSG. Students receive a grounding in computer science fundamentals of distributed computing and then work in teams as proto-typical communities in a competition to develop working integrated scientific applications, running across a set of (locally) distributed computers. In additional, several faculty now rely on OSG materials for their grid computing courses.

OSG engagement has pioneered the method of "immersive, embedded support"[5]. OSG staff, together with those from the companion NSF sponsored EIE-4CI project, spend from several days to several months focusing on a particular scientific application. They first help an individual or small sub-group run their jobs on the distributed infrastructure. They soon realise faster aggregate response and benefit from parallel execution of computations in order to produce the output to test their ideas and hypotheses. The greatest benefit is seen by those applications which can easily be parallelized, with little inter-process communication, and typically include parameter sweeps, simulation or other high throughput computing components. After initial demonstrations, viral communication and demonstration of the benefits then bring more individuals, and ultimately the new community, to the table.

Two concrete examples on the OSG in the past year are:

Dr Catherine Blake, an assistant professor in the School of Information and Library Science at the University of North Carolina, has developed a method for retrieving, analyzing and finding relationships in published research texts from multiple disciplines. "Claim Jumping through Scientific Literature" enables researchers in a given field to collect relevant published information from other fields of study. The sentence structure of millions of sentences is now generated in less than thirty hours on OSG resources, which the researchers had previously attempted on their local desktops and which would have taken several years. Analysis of individual sentences is a good application for a high-throughput facility such as OSG; the end-to-end job-run naturally carves up into smaller processing pieces. As Dr Blake told members of the OSG "reducing the processing time has enabled us to investigate otherwise prohibitively complex research questions."

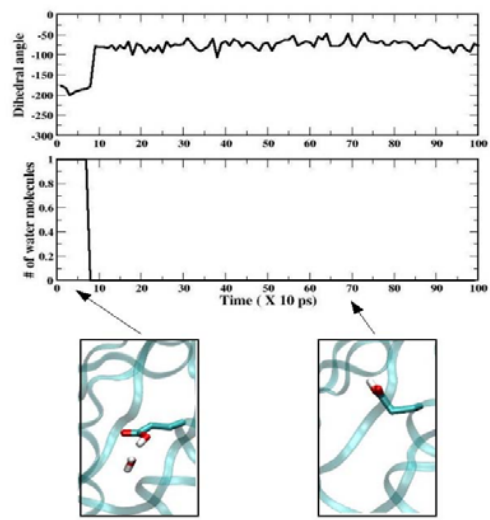

Figure 2 : Molecular dynamics simulations reveal that the staphylococcal nuclease protein residue can twist into another conformation, usually forcing water molecules to leave. But which conformation is adopted and for what proportion of the time? And what happens to the water structure around these conformations?. *Images courtesy of Johns Hopkins University*

Dr Ana Damjanovic of the National Institutes of Health and Johns Hopkins University uses CHARMM (Chemistry at HARvard Macromolecular Mechanics), a popular application developed at Harvard University for modelling the structure and behavior of molecular systems to investigate the interactions between proteins and water (Figure 2). The OSG applications support staff adapted the PANDA workflow tools, developed for the LHC ATLAS experiment, to enable short CHARMM jobs that do not need MPI, to run across OSG sites. The tools provide for automated job resubmission on failure and the submission of hundreds of jobs in a sequence. The JHU group adapted their habits and models to submit large numbers of jobs simultaneously. Simulations on OSG were used to study hydration of the side chain of the internal Glu-66 in staphylococcal nuclease. Results of the study have been submitted[6].

## 3. New Science Institutions
The OSG implements an architecture of federated infrastructures where the local-area and the wide-area are connected through locally managed interfaces. This architectural foundation allows new collaborating institutions and regional groups to locally and independently adopt OSG software and services by in a self-organized and self-managed framework. It allows the understanding of and trust in the technologies and methods to develop locally before introducing the added challenges of accessing and being accessed by the remote "anonymous" ensemble.

The large physics communities, the main scientific drivers of the OSG, have data and computing systems that span hundreds of universities worldwide. Through OSG and its peer EGEE infrastructure in Europe (and other national grid infrastructures such as NorduGrid in Scandinavia), the LHC experiments have a goal of a providing continuously operating, robust (>99% availability of) critical services and resources. Data is distributed at sustained rates of hundreds of megabytes per second[7]. More than three hundred thousand jobs are executed daily. A scientist at any site expects to retrieve and access data from any other site in the global mesh and to seamlessly run their data analysis and simulation codes at any collaborating institution that has the data available. New institutions dynamically join the facility when they are accepted into the physics collaboration and as they are ready to provide local resources and/or access and analyze the data. This is bringing about a cultural advance in the ability of university physics departments, far from the accelerator site and central computing farms and tape stores in Switzerland, to sustain active local research groups while at the same time contributing to common scientific discovery worldwide. The more than fifteen Tier-2s (experiment-funded) and ten Tier-3 (locally-funded) universities in the US that are part of the OSG infrastructure, as well as the emerging Tier-2 in South Africa, and six sites in Central and South America are such examples.

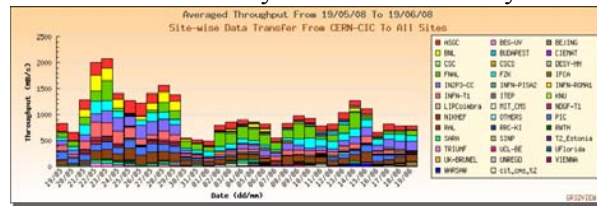

**Figure 3:** Data distribution throughput for the LHC experiments from CERN to centers worldwide. Courtesy of Ian Bird, CERN

OSG actively promotes this model through collaboration and partnerships with campus and regional infrastructures in the US and elsewhere. Each organization may adopt some, or all, of the OSG software and services as needed and provide OSG gateways for the bi-directional transport of jobs and data. The ability and reach of institutions is expanded in new ways by fostering local expertise and support as well leading to broader participation in national and international cyber-infrastructure.

The New York State Grid (NYSGrid) consortium and the Clemson University Campus Infrastructure are distinct examples that have been built up over the past twelve months. NYSGrid formed from a local consortium of universities in New York State that partners with the OSG to allow use of and support for the OSG VDT. About half of the sites directly interface to the OSG as well as participate in the local NYSGrid services and policies. The system administrators of one or two sites participate in OSG activities and then interface, educate and inform the rest. Dr Andrew Schultz, a computational chemist in the laboratory of Professor David Kofke, was engaged in running his applications on OSG via this model. Within two weeks of being introduced to OSG in Spring of 2008, he was systematically using resources and within two months was running across more than ten sites on the OSG and NYSGrid itself and with throughput of more than two hundred and fifty CPUdays per day[8].

The Clemson University Grid is a cross-campus Condor pool of Windows computers with an aggregate of more than two thousand cores. Researchers, faculty and students have agreed to share their resources locally. This local infrastructure is interfaced to the OSG via a Linux head-node which allows jobs and data to be controlled as they flow between the OSG and the Clemson grid. Resources on this ensemble of clusters were almost saturated within six months of building the environment. Researchers in industrial engineering have been among the first communities to adapt their codes and benefit from this campus-wide shared facility.

## 4. New Scientific Modalities

The OSG aims to support science at all scales – from the single PI to the largest, multi-thousand member scientific collaborations. Technologies and processes are built to support groups of individuals from the small to the large, cascading of sub-groups within these groups and, eventually, dynamic grouping and re-grouping to support ad-hoc short-term collaborations stimulated through collaborative and interdisciplinary thought and work.

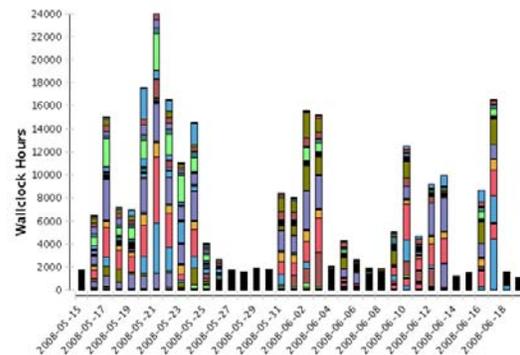

Figure 4: Cyclic use of OSG by the Engagement community

The OSG model of sharing resources across multiple communities leads to a smoothing out of the periodicity in use by each individual collaboration. This periodicity follows the natural cycle of proposing the methods and algorithms to do the research, coding and debugging of the software to implement them, production runs to obtain the outputs, and then the offline analysis, interpretation and publication of the results. The timelines of these periodicities can vary greatly. Use of the OSG is typically only significant while the production runs are being done. Figure 4 shows four such cycles of use of the OSG for the engagement community over the past six weeks. The periodicity for the large physics experiments is typically greater before the accelerator data arrives and is expected to be non-existent in general afterwards due to the overlapping of the four phases of data scientific analysis described above.

The OSG methodology of working towards an "application-ready" infrastructure and "distributed infrastructure-ready" applications allows for two things: sustained aggregate throughput of the whole facility to smooth out the cycle in usage due to these periodicities; and throughput for each community to be as large and responsive as possible when needed. The OSG model of up-front agreement on the policies for sharing, in concert with technologies to support easily changing the policies as needed, allows for dynamic use of the system without the per-event overhead of negotiation and allocation at the time of the actual need.

## 5. Summary

In summary, through its collaborative community-based activities using existing methodologies and tools, the OSG immediately benefits and stimulates new science. In addition, the activities of the OSG help evolve the organization and technical frameworks of scientific collaborations to broadly transform how collaborative data and compute intensive science will be done in the future.

**Acknowledgments**

This work was supported by the Office of Science, U.S. Department of Energy, SciDAC program under Contract DE-FC02-06ER41436 and the National Science Foundation Cooperative Agreement, PHY-0621704.

**References**
[1] Pordes R. *et al*, "The Open Science Grid", 2007 J. Phys.: Conf. Ser. 78 012057
[2] Pordes R. *et al*, "The Open Science Grid - Its Status and Implementation Architecture", to appear in *JPCS Proceedings of International Conference on Computing in High Energy and Nuclear Physics (CHEP 07)*.
[3] Bird, I et al, Deploying the LHC computing grid - the LCG service challenges; *Local to Global Data Interoperability - Challenges and Technologies, 2005*

[4] The Virtual Data Toolkit, http://vdt.cs.wisc.edu
[5] National Science Foundations Award, CI-Team Implementation Project: Embedded Immersive Engagement for Cyberinfrastructure
[6] An Open Science Grid study of the coupling between conformation and water content in the interior of a protein Ana Damjanovi et al, submitted to the Journal of Physical Chemistry, B, April 2008.
[7] Schulz M. *et al.*, "Building the WLCG file transfer service", to appear in *JPCS Proceedings of International Conference on Computing in High Energy and Nuclear Physics (CHEP 07)*.
[8] See publication K. M. Benjamin, A. J. Schultz and D. A. Kofke, "Virial coefficients of polarizable water: Applications to thermodynamic properties and molecular clustering", J. Phys. Chem. C 111 16021 (2007), for a description of the research being done.